\documentclass[prl,a4paper,superscriptaddress, twocolumn, showpacs,amsmath,amssymb,floatfix,amstex]{revtex4}

\usepackage{psfrag}

\usepackage[dvips]{graphicx}
\usepackage{bm,pstricks}

\input{epsf}

\begin{document}

\title{Spectral properties of the Google matrix of the World Wide Web and other directed networks}
\author{Bertrand Georgeot}
%\homepage[]{http://www.quantware.ups-tlse.fr}
\affiliation{\mbox{Laboratoire de Physique Th\'eorique (IRSAMC), 
Universit\'e de Toulouse, UPS, F-31062 Toulouse, France}}
\affiliation{\mbox{LPT (IRSAMC), CNRS, F-31062 Toulouse, France}}
\author{Olivier Giraud\footnote{present address: Laboratoire de Physique Th\'eorique et Mod\`eles Statistiques, UMR 8626 du CNRS, Universit\'e Paris-Sud, Orsay, France.} }
%\homepage[]{http://www.quantware.ups-tlse.fr}
\affiliation{\mbox{Laboratoire de Physique Th\'eorique (IRSAMC), 
Universit\'e de Toulouse, UPS, F-31062 Toulouse, France}}
\affiliation{\mbox{LPT (IRSAMC), CNRS, F-31062 Toulouse, France}}
\author{Dima L. Shepelyansky}
%\homepage[]{http://www.quantware.ups-tlse.fr}
\affiliation{\mbox{Laboratoire de Physique Th\'eorique (IRSAMC), 
Universit\'e de Toulouse, UPS, F-31062 Toulouse, France}}
\affiliation{\mbox{LPT (IRSAMC), CNRS, F-31062 Toulouse, France}}

%\date{\today}
\date{February 17, 2010}

\begin{abstract}
We study numerically the spectrum and eigenstate
properties of the Google matrix of various examples of directed networks
such as vocabulary networks of dictionaries
and university World Wide Web networks.  The spectra have gapless
structure in the vicinity of the maximal eigenvalue for Google 
damping parameter $\alpha$ equal to unity.  The vocabulary 
networks have relatively homogeneous spectral density, while
university networks have pronounced spectral structures
which change from one university to another, reflecting specific
properties of the networks.  We also determine specific properties
of eigenstates of the Google matrix, including the PageRank.
The fidelity of the PageRank is proposed as a new characterization of its
stability.

\end{abstract}

%89.20.Hh  	World Wide Web, Internet
%89.75.Hc       Networks and genealogical trees 
%05.40.Fb       Random walks and Levy flights
%72.15.Rn Localization effects (Anderson or weak localization) 
%87.23.Ge       Dynamics of social systems
% 64.60.Fr      Equilibrium properties near critical points, critical exponents

\maketitle

\section{I Introduction}

The rapid growth of the World Wide Web (WWW) brings the challenge of 
information retrieval from this enormous database which at present contains
about $10^{11}$ webpages. An efficient algorithm for classification of
webpages was proposed in \cite{brin}, and is now known as the PageRank Algorithm (PRA).
This PRA formed the basis of the Google search engine, which is used by
the majority of Internet users in everyday life.  The PRA allows to determine
efficiently a vector ranking the nodes of a network by order of importance.
This PageRank vector is obtained as an eigenvector of the Google matrix ${\bf G }$
built on
the basis of the directed links between WWW nodes (see e.g. \cite{googlebook}):
\begin{equation}
{\bf G}=\alpha {\bf S}+(1-\alpha) {\bf E}/N \; .
\label{eq1}
\end{equation}
Here ${\bf S}$ is the matrix constructed from the adjacency matrix $A_{ij}$
of the directed links of the network of size $N$, with $A_{ij}=1$ if there is a link
from node $j$ to node $i$, and $A_{ij}=0$ otherwise. Namely, 
 $S_{ij}=A_{ij}/\sum_k A_{kj}$ if $\sum_k A_{kj} > 0$, and
 $S_{ij}=1/N$ if all elements in the column $j$ of ${\bf  A}$ are zero.
The last term in Eq.(\ref{eq1}) with uniform matrix $E_{ij}=1$ describes the probability
$1-\alpha$ of a random surfer propagating along the network to jump
randomly to any other node.  The matrix ${\bf G }$ belongs to the class of
Perron-Frobenius operators.  For $0 < \alpha < 1$ it has a unique maximal eigenvalue
at $\lambda=1$, separated from the others by a gap of size at least $1-\alpha$ 
(see e.g. \cite{googlebook}).  The eigenvector associated to this maximal eigenvalue
is the PageRank vector, which can be viewed as the steady-state distribution for
the random surfer.  
Usual WWW networks correspond to very sparse matrix ${\bf A}$ and repeated applications
of ${\bf G}$ on a random vector converges quickly to the PageRank vector, after
$50-100$ iterations for $\alpha=0.85$ which is the most commonly used value \cite{googlebook}.
The PageRank vector is real nonnegative and can be ordered 
by decreasing values $p_j$, giving the relative importance
of the node $j$.  It is known that when $\alpha$ varies, all eigenvalues evolve
as $\alpha \lambda_i$ where $\lambda_i$ are the eigenvalues for $\alpha=1$ and
$i=2,...N$, while the largest eigenvalue $\lambda_1=1$, associated with the PageRank,
remains unchanged \cite{googlebook}.  

The properties of the PageRank vector for WWW have been extensively studied by the computer
science community and many important properties have been established 
\cite{donato,boldi,avrach1,avrach2,avrach3}.  For example, it was shown that
$p_j$ decreases approximately in an algebraic way $p_j \sim 1/j^{\beta}$ 
with the exponent $\beta \approx 0.9$ \cite{donato}.  It is also known that typically for 
the Google matrix of WWW
at $\alpha=1$ there are many eigenvalues very close or equal to $\lambda=1$, and that
even at finite $\alpha <1$ there are degeneracies of eigenvalues with $\lambda=\alpha$
(see e.g. \cite{capizzano}).  

In spite of the important progress obtained during these investigations of PageRank vectors,
the spectrum of the Google matrix ${\bf G}$ was rarely studied as a whole.  Nevertheless,
it is clear that the structure of the network is directly linked to this spectrum.  
Eigenvectors other than the PageRank describe the relaxation processes toward
the steady-state, and also characterize various communities or subsets of the network.
Even if models of directed networks of small-world type \cite{dorogovtsev}
have been analyzed, constructed and investigated,
the spectral properties of matrices corresponding to such networks were not so much studied.
Generally for a directed network the matrix ${\bf G}$ is nonsymmetric and thus the
spectrum of eigenvalues is complex.
Recently the spectral study of the Google matrix for the Albert-Barabasi (AB) model \cite{albert}
and randomized university WWW networks was performed in \cite{ggs}.  For the AB model
the distribution of links is typical of scale-free networks \cite{dorogovtsev}.  The distribution
of links for the university network is approximately the same and is not affected
by the randomization procedure.  
Indeed, the randomization procedure corresponds to the one proposed in
\cite{maslov2} and is performed by taking pairs  of links and inverting the initial vertices,
keeping unchanged the number of ingoing and outgoing links for each vertex. 
It was established that the spectra of the AB model and the randomized university
networks were quite similar.   Both have a large gap between the largest eigenvalue
$\lambda_1=1$ and the next one with $|\lambda_2|\approx 0.5$ at $\alpha=1$.  This is in
contrast with the known property of WWW where $\lambda_2$ is usually very close or equal to 
unity \cite{googlebook,capizzano}.  Thus it appears that the AB model and the
randomized scale-free networks have a very different spectral structure compared to
real WWW networks. Therefore it is important to study the spectral
properties of examples of real networks (without randomization).  

In this paper, we thus study the spectra of Google matrices for the WWW networks of several
universities and show that indeed they display very different properties compared
to random scale-free networks considered in \cite{ggs}.  We also explore the spectra
of a completely different type of real network, built from the vocabulary links
in various dictionaries.  In addition, we analyze the properties of eigenvectors of
the Google matrix for these networks.
A special attention is paid to the PageRank 
vector and in particular we characterize its sensitivity to $\alpha$ through a new quantity, the 
PageRank fidelity.

The paper is organized as follows.  In Section II we give the description of
the university and vocabulary networks whose Google matrices we consider.
The properties of spectra and eigenstates are investigated in Section III.
The fidelity of PageRank and its other properties are analyzed in Section IV.
Section V explores various models of random networks for which the spectrum can
be closer to the one of real networks. The conclusion is given in Section VI. 

\section{II Description of networks of university WWW and dictionaries} 

In order to study the spectra and eigenvectors of Google matrices
of real networks, we numerically explored several systems.

Our first example consists in the WWW networks of UK universities,
taken from the database \cite{uni}. The vertices are the
HTML pages of the university websites in 2002. The links correspond
to hyperlinks in the pages directing to another webpage.
To reduce the size of the matrices in order to perform exact diagonalization, 
only webpages with at least
one outlink were considered.  There are still dangling nodes,
despite of this selection, since some sites have outlinks only
to sites with no outlink.  We checked on several examples
that the general properties of the spectra were not affected
by this reduction in size.  We present data on the spectra
from five universities:
\begin{itemize}
\item  University of Wales at Cardiff (www.uwic.ac.uk), with 2778 sites
  and 29281 links.
\item  Birmingham City University (www.uce.ac.uk); 10631 sites and 82574 links.
\item  Keele University (Staffordshire) (www.keele.ac.uk); 11437 sites and 67761 links.
\item  Nottingham Trent University (www.ntu.ac.uk); 12660 sites and 85826 links.
\item   Liverpool John Moores University (www.livjm.ac.uk); 13578 sites and 111648 links.
\end{itemize}

A much larger sample of university networks from the same database was
actually used, including universities from the US, Australia and New Zealand,
in order to insure that the results presented were representative.

As opposed to the full spectrum of the Google matrix, the PageRank
can be computed and studied for much larger matrix sizes.  In the
studies of Section IV, we therefore included additional data
from the university networks of Oxford in 2006 (www.oxford.ac.uk)
with 173733 sites and 2917014 links
taken from \cite{uni}, and the network of Notre Dame University from the US 
taken from the database \cite{laszlo} with 325729 sites and 1497135 links (without removing any node).

In addition, we also investigated several vocabulary networks constructed
from dictionaries; the network data were taken from \cite{data-dicos}.
\begin{itemize}
\item Roget dictionary (1022 vertices and 5075 links) \cite{roget}.\\
The 1022 vertices correspond to the categories in the
1879 edition of Roget's Thesaurus of English Words and Phrases.
There is a link from category X to category Y if
Roget gave a reference to Y among the words and phrases of X, 
or if the two categories are related by
their positions in the book.

\item ODLIS dictionary (Online Dictionary of Library and Information
  Science), version December 2000 \cite{odlis} (2909 vertices and 18419 links).\\
A link (X,Y) from term X to term Y is created if the term Y is used in the definition of term X. 
 
\item FOLDOC dictionary (Free On-Line Dictionary of Computing) \cite{foldoc}
(13356 vertices and 120238 links)\\
A link (X,Y) from term X to term Y is created if the term Y is used in the definition of term X. 

\end{itemize}

Distribution of ingoing and outgoing links for these university WWW
networks is similar to those of much larger WWW networks discussed
in \cite{donato,avrach3,dorogovtsev}.  An example is shown
in the Appendix for the network
of Liverpool John Moores University, together with data from
AB models discussed in \cite{ggs} (see Fig. \ref{fig12}).

\section{III Properties of spectrum and eigenstates}

To study the spectrum of the networks described in the previous section, we
construct the Google matrix ${\bf G}$ associated to them at $\alpha=1$.  
After that 
the spectrum $\lambda_i$ and right eigenstates $\psi_i$ 
 of ${\bf G}$ (satisfying the relation ${\bf G}\psi_i =\lambda_i \psi_i$)
 are computed by direct
diagonalization using standard LAPACK routines.  Since ${\bf G}$ is 
generally a nonsymmetric matrix for our networks, the eigenvalues $\lambda_i$
are distributed in the complex plane, inside the unit disk $|\lambda_i|\leq 1$.

\begin{figure}[htbp]
\begin{center} 
\includegraphics[width=.95\linewidth]{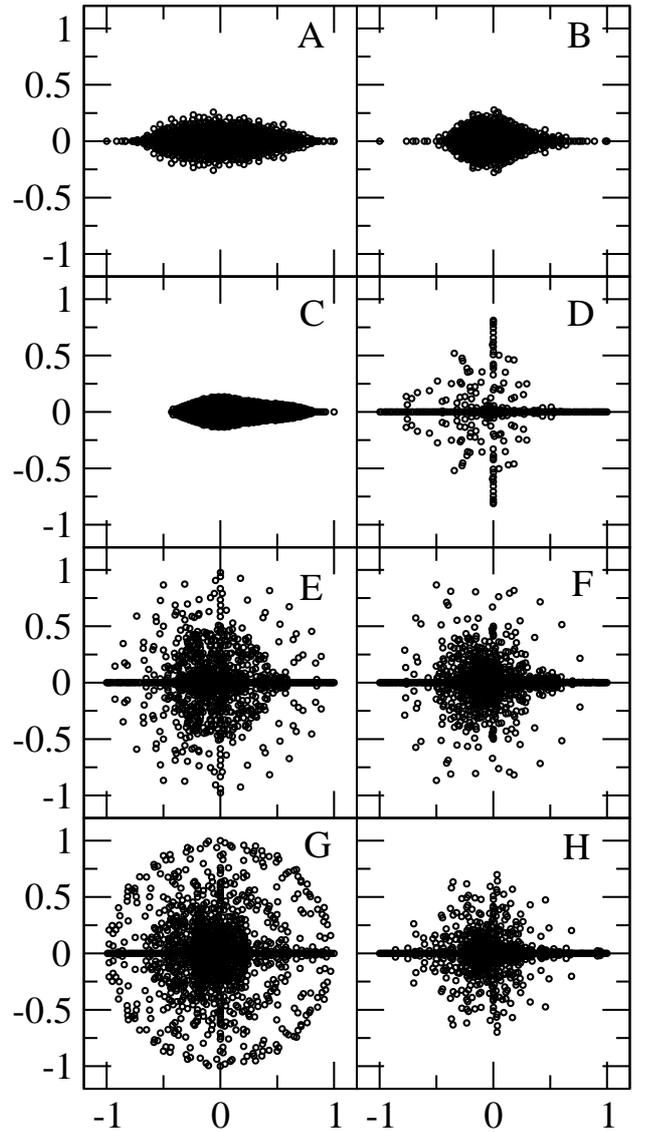}
\end{center} 
\vglue +0.5cm
\caption{Distribution of eigenvalues $\lambda_i$ of Google
  matrices in the complex plane at $\alpha=1$ for dictionary networks:
Roget (A, N=1022), ODLIS
  (B, N=2909) and FOLDOC (C, N=13356); university WWW networks:
University of Wales (Cardiff) (D,
  N=2778), Birmingham City University (E, N=10631),
 Keele University (Staffordshire)  (F, N=11437), 
Nottingham Trent University (G, N=12660),   Liverpool John Moores University
  (H, N=13578)(data for universities are for 2002).}
 \vspace{2cm}
\label{fig1}
\end{figure}

The spectrum for our eight networks is shown in Fig.\ref{fig1}.  An important
property of these spectra is the presence of eigenvalues very close to 
$\lambda=1$ and moreover we find that $\lambda=1$ eigenvalue has
significant degeneracy.  It is known that such an exact degeneracy is 
typical for WWW networks (see e.g.  \cite{avrach3,capizzano}).  In addition to
this exact degeneracy, there are quasidegenerate eigenvalues very close
to $\lambda=1$.  It is important to note that these features are absent
in the spectra of random networks studied in \cite{ggs} based on
the AB model and on the randomization of WWW university networks, where the
spectrum is characterized by a large gap between the first eigenvalue
$\lambda_1=1$ and the second one with $|\lambda_2| \approx 0.5$. 
For example, the spectrum shown in Fig.\ref{fig1} panel H corresponds
to the same university whose randomized spectrum was displayed 
in Fig.1 (bottom panel) in \cite{ggs}.  Clearly the structure 
of the spectrum becomes very different after randomization of links.
Another property of the spectra displayed in Fig.\ref{fig1} that we want 
to stress is the presence
of clearly pronounced structures which are different from one network to
another.  The structure is less pronounced in the case of
the three spectra obtained from dictionary networks.  In this case, 
the spectrum is flattened, being closer to the real axis. In contrast, for
the WWW university networks, the spectrum is spread out over the unit disk.
However, there is still a significant fraction of eigenvalues 
close to the real axis.  We understand this feature by the existence 
of a significant number of symmetric ingoing and outgoing links
(48 \% in the case of the Liverpool John Moores University network),
which is larger compared to the case of randomized university networks
considered in \cite{ggs}.

\begin{figure}[htbp]
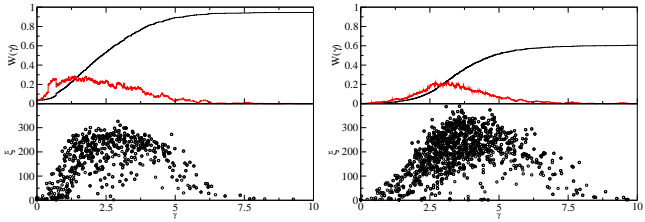

\begin{center} 
\includegraphics[width=.48\linewidth]{fig2a.eps}
\includegraphics[width=.48\linewidth]{fig2b.eps}
\end{center} 
\caption{(Color online) Left: Roget dictionary, $\alpha=1$.
Top panel: normalized density of states $W$ (black) obtained as a 
  derivative of a smoothed version of the integrated
  density (smoothed over a small interval $\Delta\gamma$ 
varying with matrix size),
  integrated density  is shown in red (grey) . Bottom panel: PAR of eigenvectors as a function
  of $\gamma$; degeneracy of $\lambda=1$ is 18 (note that the value $W(0)$
  corresponds to eigenvalues with  $|\lambda|=1$). Right: ODLIS dictionary, same as left; 
degeneracy of $\lambda=1$ is 4.}
\label{fig2}
\end{figure}

\begin{figure}[htbp]
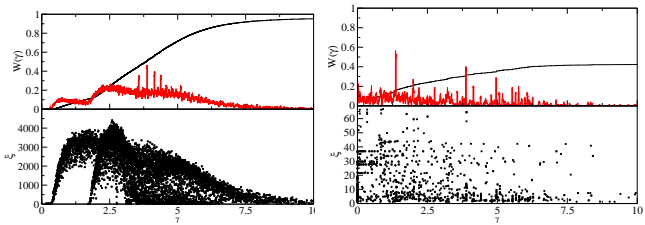

\begin{center} 
\includegraphics[width=.48\linewidth]{fig3a.eps}
\includegraphics[width=.48\linewidth]{fig3b.eps}
\end{center} 
\vspace{1cm}
\caption{(Color online) Left: FOLDOC dictionary, same as Fig.~\ref{fig2}; 
degeneracy of $\lambda=1$ is 1; Right: University of Wales (Cardiff), 
same as left; degeneracy of $\lambda=1$ is 69. }
\label{fig3}
\end{figure}

\begin{figure}[htbp]
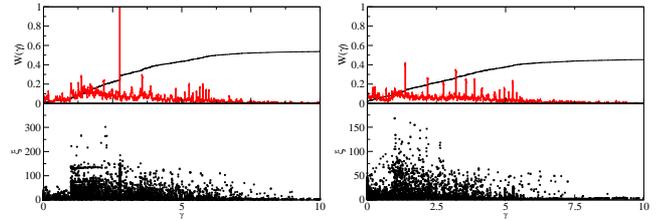

\vglue 0.3cm
\begin{center} 
\includegraphics[width=.48\linewidth]{fig4a.eps}
\includegraphics[width=.48\linewidth]{fig4b.eps}
\end{center} 
\vspace{1cm}
\caption{(Color online) Left: Birmingham City University, same as Fig.~\ref{fig2}; 
degeneracy of $\lambda=1$ is 71; Right: Keele University (Staffordshire), 
same as left; degeneracy of $\lambda=1$ is 205. }
\label{fig4}
\end{figure}

\begin{figure}[htbp]
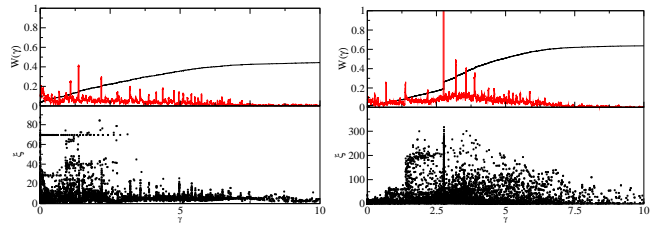

\vglue 0.3cm
\begin{center} 
\includegraphics[width=.48\linewidth]{fig5a.eps}
\includegraphics[width=.48\linewidth]{fig5b.eps}
\end{center} 
\vspace{1cm}
\caption{(Color online) Left: Nottingham Trent University, same as Fig.~\ref{fig2}; 
degeneracy of $\lambda=1$ is 229. Right: Liverpool John Moores University, 
same as left; degeneracy of $\lambda=1$ is 109; other
degeneracy peaks correspond to $\lambda=1/2$ (16), $\lambda=1/3$ (8); 
$\lambda=1/4$ (947), $\lambda=1/5$ (97), being located at 
$\gamma=-2 \ln \lambda$; other degeneracies are also present, e.g. 
$\lambda=1/\sqrt{2}$ (41).
}
\label{fig5}
\end{figure}

To characterize the spectrum, we introduce the relaxation rate
$\gamma$ defined by the relation $|\lambda|= \exp (-\gamma/2)$. 
For characterization of eigenvectors $\psi_i(j)$, 
we use the PArticipation Ratio
(PAR) defined by $\xi =(\sum_j |\psi_i(j)|^2)^2/\sum_j |\psi_i(j)|^4$.
This quantity gives the effective number of vertices of the network
contributing to a given eigenstate $\psi_i$; it is often used 
in solid-state systems with disorder to characterize localization properties
(see e.g. \cite{mirlin}).  The dependence of the density of states $W(\gamma)$
in $\gamma$, which gives the number of eigenstates in the interval
$[\gamma, \gamma+d\gamma]$, is shown in Figs.\ref{fig2},\ref{fig3},\ref{fig4},\ref{fig5} (top panels).
The normalization is chosen such that $\int_0^{\infty} W(\gamma)d\gamma=1$,
corresponding to the total number of eigenvalues $N$ 
(equal to the matrix size).  We also show the integrated version of 
this quantity in the same panels.
In the same Figs we show the PAR $\xi$ of the eigenstates as a function of $\gamma$ (bottom panels).

It is clear that for the dictionary networks the density of states $W$
depends on $\gamma$ in a relatively smooth way, with a broad maximum at 
$\gamma\approx 1-2$.  The distribution of PAR has also
a maximum at approximately the same values.
The case of the dictionary FOLDOC is a bit special,
showing a bimodal distribution which is also clearly seen in the dependence
of $\xi$ on $\gamma$.  This comes from the fact that the distribution
of eigenvalues in Fig. \ref{fig1} (panel C) 
is highly asymmetric with respect to the imaginary axis. 
The latter case has also no degeneracy at $\lambda=1$.
In these three networks the density of states decreases for $\gamma$
approaching $0$.  We note that the integrated version of the density of states
reaches a plateau for $\gamma \geq 6-7$. This saturation value is less 
than $1$, meaning that a certain nonzero fraction of eigenvalues
are extremely close to $\lambda=0$.

For the WWW university networks, the density of states is much more 
inhomogeneous in $\gamma$. Even if a broad maximum is visible, there are
sharp peaks at certain values of $\gamma$.  The sharpest peaks correspond
to exact degeneracies at certain complex values of $\lambda$.  The
degeneracies are especially visible at the real values $\lambda=1/2,
\lambda=1/3$ and other $1/n$ with integer values of $n$.  We attribute 
this phenomenon to the fact that the small number of links
gives only a small number of different values for the matrix elements of
the matrix ${\bf G}$.  For the university networks, 
the degeneracy at $\lambda=1$ is much larger than in the case of dictionaries.
The integrated densities of states show visible vertical jumps which
correspond to the degeneracies; their growth saturates at $\gamma\approx 7$
showing that about $30-50 \%$ of the eigenvalues are located in the vicinity
of $\lambda=0$.

The PAR distribution for the university networks fluctuates strongly,
even if a broad maximum is visible.  Typical values have $\xi\approx 100$,
which is small compared to the matrix size $N\sim 10^4$.  This indicates
that the majority of eigenstates are localized on certain zones of the
network.  This does not exclude that certain eigenstates with 
a larger $\xi$ will be delocalized on a large fraction of the network
in the limit of very large $N$.

\begin{figure}[htbp]
\vglue 0.8cm
\begin{center} 
\includegraphics[width=.95\linewidth]{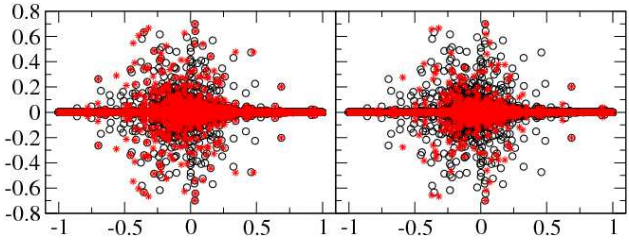}
\end{center} 
\vspace{1cm}
\caption{Cloud of eigenvalues for Liverpool John Moores University, $\alpha=1$.
Circles: full matrix $N=13578$. Stars:  truncated matrix of size $8192$
(left) and $4096$ (right).}
\label{fig6}
\end{figure}

The exact ${\bf G}$ matrix diagonalization requires significant computer memory and
is practically restricted to matrix size $N$ of about $N < 30000$.  However, real networks
such as WWW networks can be much larger.  It is therefore important to find numerical 
approaches in order to obtain the spectrum of large networks using approximate methods.
A natural possibility is to order the sites through the PageRank method and to consider
the spectrum of the (properly renormalized)
truncated matrix restricted to the sites with PageRank larger
than a certain value.  In this way, the truncation takes into account the most
important sites of the network.  The effect of such a truncation
is shown in Fig. \ref{fig6} for the largest network of our sample.  The numerical data
show that the global features of the spectrum are preserved by moderate truncation,
but individual eigenvalues deviate from their exact values when more
than $50$\% of sites are truncated. Probably future developments of this approach 
are needed in order to be able to truncate a larger fraction of sites.

\section{IV Fidelity of PageRank and its other properties}

In the previous section we studied the properties of the full spectrum and all
eigenstates of the ${\bf G}$ matrix for several real networks.  The PageRank is 
especially important since it allows to obtain an efficient classification
of the sites of the network \cite{brin,googlebook}.  Since the networks
usually have small number of links, it is possible to obtain the
PageRank by vector iteration for enormously large size of
networks as described in \cite{brin,googlebook}.

\begin{figure}[htbp]
\vglue 0.5cm
\begin{center} 
\includegraphics[width=.95\linewidth]{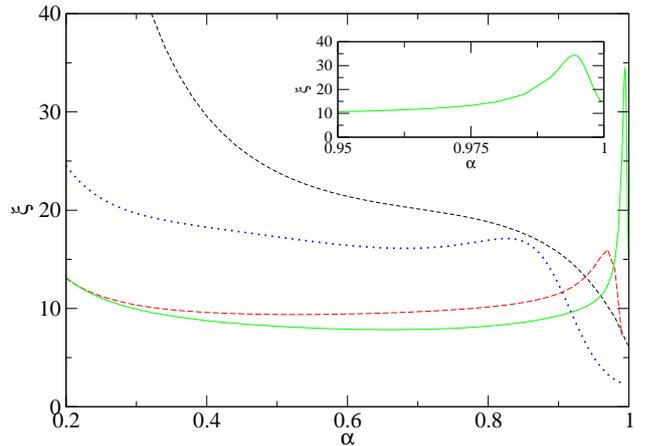}
\end{center} 
\vspace{1cm}
\caption{(Color online) PAR $\xi$ of PageRank as a function of $\alpha$ 
 for University of Wales (Cardiff)
(black/dashed), Notre-Dame
  (blue, dotted), Liverpool John Moores University (red/long dashed) and Oxford (green/solid)  
Universities (curves from top to bottom at $\alpha=0.6$). Network sizes vary
from $N=2778$ to $N=325729$. Inset is a zoom
for data from Oxford University close to $\alpha=1$.}
\label{fig7}
\end{figure}

Due to this significance of the PageRank, it is important to characterize its properties.
In addition, it is important to know how sensitive the PageRank is with respect to
the Google parameter (damping parameter) $\alpha$ in Eq.~(\ref{eq1}). The localization
property of the PageRank can be quantified through the PAR $\xi$ defined above.
The dependence of $\xi$ on $\alpha$ is shown in Fig.\ref{fig7} for four University WWW
networks, including two from Fig.\ref{fig1} (panels D and H) and two of much larger
sizes (Notre Dame and Oxford).  For $\alpha \rightarrow 0$ the PAR goes to the matrix size
since the ${\bf G}$ matrix is dominated by the second part of Eq.~\ref{eq1}.  However,
in the interval  $0.4 < \alpha< 0.9 $ the dependence on $\alpha$ is rather weak, indicating
stability of the PageRank.  For $0.9<\alpha < 1$ the PAR value has a local maximum where its
value can be increased by a factor $2-3$.  We attribute this effect to the existence of
an exact degeneracy of the eigenvalue $\lambda=1$ at $\alpha=1$, discussed in the previous section.
In spite of this interesting behavior of $\xi$ in the vicinity of $\alpha=1$, the
value of $\xi$, which gives the effective number of populated sites, remains much smaller
than the network size.  In other models considered in \cite{ggs,zs}, a delocalization
of the PageRank was observed for some $\alpha$ values,  so that $\xi$ was growing 
with system size $N$. For the WWW university networks considered here, delocalization
is clearly absent (network sizes in Fig. \ref{fig7} vary over two order of magnitudes).  
This is in agreement with the value of the exponent $\beta \approx 0.9$ for
the PageRank decay, which was found for large samples of WWW data in \cite{donato,avrach3}.
Indeed, for that value of $\beta$, PAR should be independent of system size.

\begin{figure}[htbp]
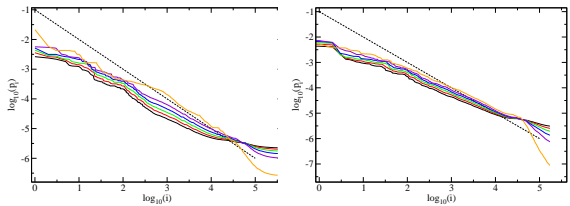

\vglue 0.3cm
\begin{center} 
\includegraphics[width=.42\linewidth]{fig8a.eps}
\includegraphics[width=.42\linewidth]{fig8b.eps}
\end{center} 
\vspace{1cm}
\caption{(Color online) Some PageRank vectors $p_j$ for Notre-Dame university
  (left panel) and Oxford (right panel). From top to bottom at $\log_{10} (i)=5$:
  $\alpha$=0.49 (black), 0.59 (red), 0.69 (green), 0.79 (blue) , 0.89
  (violet) and 0.99 (orange). Dashed line indicates the slope -1.}
\label{fig8}
\end{figure}

Our data for PageRank distribution also show its stability
as a whole for variation of $\alpha$ in the interval $0.4 < \alpha< 0.9 $,
as it is shown in Fig. \ref{fig8}.

\begin{figure}[htbp]
\begin{center} 
\includegraphics[width=.48\linewidth]{fig9a.eps}
\includegraphics[width=.48\linewidth]{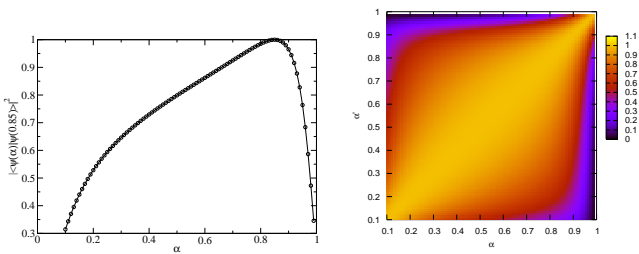}
\end{center} 
\vspace{1cm}
\caption{(Color online) PageRank fidelity $f(\alpha,\alpha')$ for Notre-Dame university ($N=325729$); 
left panel: $f(\alpha,\alpha'=0.85)=|\langle\psi(\alpha)|\psi(0.85)\rangle|^2 $ (see
Eq.~(\ref{fidelity})); right panel: color density plot of $f(\alpha,\alpha')$ .}
\label{fig9}
\end{figure}

The sensitivity of the PageRank with respect to $\alpha$ can be more precisely characterized
through the {\em PageRank fidelity} defined as

\begin{equation}
f(\alpha,\alpha')=|\sum_j \psi_1(j,\alpha)\psi_1(j,\alpha')|^2 \; ,
\label{fidelity}
\end{equation}
where $\psi_1(j,\alpha)$ 
is the eigenstate at $\lambda=1$ of the Google matrix ${\bf G}$ with parameter $\alpha$ in
Eq.~(\ref{eq1}); here the sum over $j$ runs over the network sites (without PageRank reordering).
We remind that the eigenvector $\psi_1(j,\alpha)$  is normalized by $\sum_j \psi_1(j,\alpha)^2=1$.
Fidelity is often used in the context of quantum chaos and quantum computing to characterize
the sensitivity of wavefunctions with respect to a perturbation \cite{prosen,klaus}.  The variation
of this quantity with $\alpha$ and $\alpha'$ is shown in Fig. \ref{fig9}.  The fidelity reaches its
maximum value $f=1$ for $\alpha=\alpha'$.  According to Fig. \ref{fig9} (right panel), the stability
plateau where fidelity
remains close to 1, indicating stability of PageRank, is broadest for $\alpha=0.5$.  This is in
agreement with previous results presented in  \cite{avrach4}, where the
same optimal value of $\alpha$ was found based on different arguments.

\section{V Spectrum of model systems}

\begin{figure}[htbp]
\vspace{1cm}
\begin{center} 
\includegraphics[width=.95\linewidth]{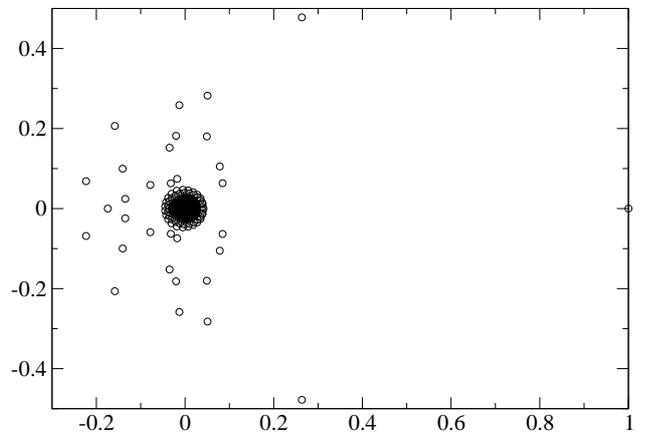}
\end{center} 
\vspace{1cm}
\caption{Spectrum of eigenvalues $\lambda$ in the complex plane
for the Avrachenkov-Lebedev model
of \cite{avrach1}, with
 $N=2^{11}$ (network size), 
$\alpha=0.85$, $m=5$ outgoing links per node.  
Multiplicity of links is taken into account
in the construction of ${\bf G}$.
%Note that there can be several links
%  from $i$ to $j$; the entry $S_{ji}$ takes into account this multiplicity
%  (i.e. it is not just the normalized adjacency matrix).
\label{fig10}}
\end{figure}

The results obtained in \cite{ggs} compared to those presented in the previous
section show that while the spectrum of the network has a large gap
between $\lambda=1$ and the other eigenvalues, still
certain properties of the PageRank can be similar in both cases (e.g.
the exponent $\beta$).  In fact the studies performed in
the computer science community were often based on simplified
models, which can nevertheless give the value of $\beta$ close
to the one of real networks.  For example, the model studied by
Avrachenkov and Lebedev \cite{avrach1} allows to obtain analytical
expressions for $\beta$ with a value close to the one obtained  
for WWW.  It is interesting to see what are the spectral properties
of this model.  In Fig. \ref{fig10} we show the spectrum for this
model for $\alpha=0.85$.  
Our data show that this model has an enormous gap, thus 
being very different from 
spectra of real networks shown in Fig. \ref{fig1}.

The above results, together with those of 
\cite{ggs}, show that many commonly used network models 
are characterized by a large gap between $\lambda=1$ and the
second eigenvalue, in contrast with real networks. In order to build a network model
where this gap is absent, we introduce here what we
call the color model.  It
 is an extension of the AB model, that allows to obtain
results for the spectral distribution that are closer to real 
networks. We divide the nodes into  $n$ sets (''colors''), allowing $n$ to grow
with network size.
Each node is labeled by an integer between 0 and $n-1$.
 At each
step, links and nodes 
are added as in the AB model but also
with probability $\eta$ the new node is
  introduced with a new color. The only links authorized between nodes are links 
within each set.
Such a structure implies that the second eigenvalue of matrix $G$
is real and exactly equal to $\alpha$ \cite{kamvar}. The colors
correspond to communities in the network.

In order to have a more realistic model, we allow for the rule
for links to be broken with some probability $\varepsilon$. That is, 
at each time step an link between two nodes is chosen at
random according to the rules of the AB model. Then if it
agrees with the color rule above it is used; if it
does not then with probability  $1-\varepsilon$
it is just omitted, and with probability $\varepsilon$ it is nevertheless 
added.

\begin{figure}[htbp]
\vglue 0.8cm
\begin{center} 
\includegraphics[width=.95\linewidth]{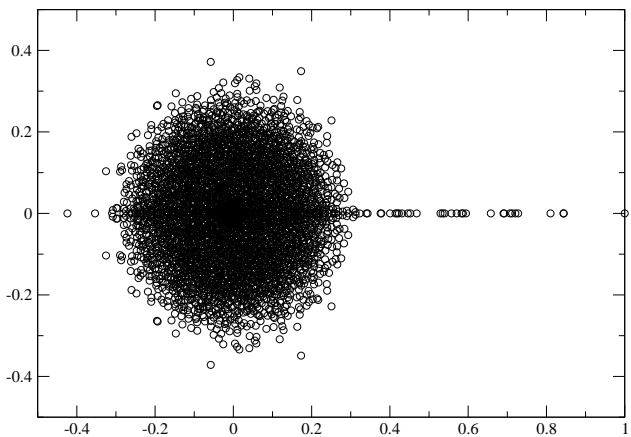}
\end{center} 
\vspace{1cm}
\caption{Spectrum of eigenvalues $\lambda$ 
in the complex plane for the color model, $N=2^{13}$, $p=0.2$,
  $q=0.1$, $\alpha=0.85$. 
Nodes are divided into $n$ color sets labeled from $i=0$ to $n-1$;
nodes and links are created according to AB model; 
only authorized links  are links within a color set  $i$.
This rule is broken with probability $\epsilon=10^{-3}$.
We start with 3 color sets; with probability $\eta$ a new color is
  introduced (we take $\eta=10^{-2}$). In the example displayed, when the number of nodes
  reaches $N$, $n=83$ colors.\label{fig11}}
\end{figure}

\begin{figure}[!h]
\vglue 0.8cm
\begin{center} 
\includegraphics[width=.95\linewidth]{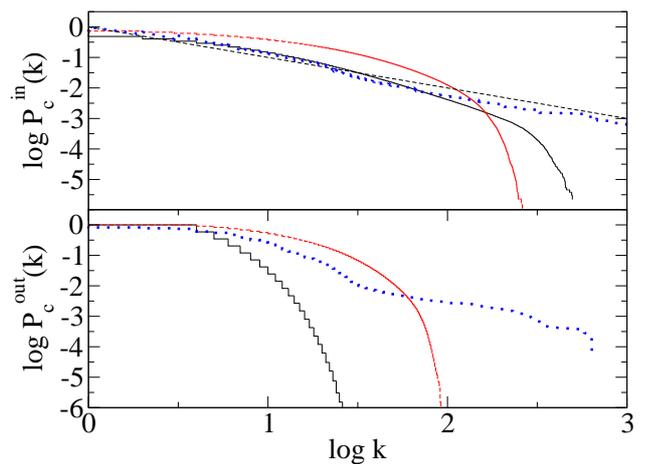}
\caption{Cumulative distribution of ingoing links 
$P_c^{in}(k)$ (top panel) and of outgoing links $P_c^{out}(k)$ 
(bottom panel) 
for the AB model with
vector size $N=2^{14}$, for $q=0.1$ (black/solid) and $q=0.7$ (red/dashed), 
data are averaged over 80 realizations of AB model, and for the network 
of Liverpool John Moores  University with $N=13578$, 
(panel H in Fig.~\ref{fig1}) (blue/dotted).
Average number of in- or outgoing links is $<k>=6.43$ for $q=0.1$,
$<k>=14.98$ for $q=0.7$, $<k>=8.2227$ for LJMU. Dashed straight line
indicates the slope -1. Logarithms are decimal. 
%Stair-like behavior of the
%curves corresponds to round-off errors.
}
\label{fig12}
\end{center} 
\end{figure}

The spectrum of this color model is shown
in Fig. \ref{fig11} for $\alpha=0.85$.  The second eigenvalue
is now exactly at $\lambda=0.85$, demonstrating the absence of a gap.
There is also a set of eigenvalues which is located on the real
line, but the majority of states remains inside a circle
$|\lambda| < 0.3$ as in the AB model.  

Thus the color model allows to eliminate the gap, but still the 
distribution of eigenvalues $\lambda$ in the complex plane remains 
different from the spectra of real networks shown in Fig. \ref{fig1}:
the structures prominent in real networks are not visible, and
eigenvalues in the gap are concentrated only on
the real axis or close to it. 

\section{VI Conclusion} 

In this work we performed numerical analysis of the spectra and
eigenstates of the Google matrix ${\bf G}$ for several real networks.
The spectra of the analyzed networks have no gap between first and second
eigenvalues, in contrast with commonly used scale-free network models (e.g. AB model).
The spectra of university WWW networks are characterized by complex structures
which are different from one university to another.  At the same time, 
PageRank of these university networks look rather similar.  In contrast, the
Google matrices of vocabulary networks of dictionaries have spectra with
 much less structure.  

These studies show that usual models of random scale-free networks
miss many important features of real networks.  In particular,
they are characterized by a large spectral gap, which is generally absent
in real networks.  We attribute the physical origin of
this gap to the known property of small-world and scale-free networks 
that only logarithmic time (in system size)
is needed to go from any node to any other node 
(see e.g. \cite{dorogovtsev}).  Due to that, the relaxation process
in such networks is fast and the gap, being inversely
proportional to this time, is accordingly very large.  In contrast,
the presence of weakly coupled communities in real networks
makes the relaxation time very large, at least for certain configurations.
Therefore, it is desirable to construct new random scale-free models which
could capture in a better way the actual properties of real networks.
The color model presented here is a first step in this direction.
We note that Ulam networks built from dynamical maps can capture 
certain properties of real networks in a relatively good manner 
\cite{zs,ermann}.  In these latter networks, it is possible to
have a delocalization of the PageRank when $\alpha$ or map
parameters vary; we didn't observe such feature here.  

Indeed, our data show that the PageRank remains localized for 
all values of $\alpha >0.3$. 
We also showed that the use of the fidelity as a new quantity 
to characterize the stability of PageRank enables to identify
a stability plateau located around $\alpha=0.5$.  

We think that future investigation of the spectral properties 
of the Google matrix will open new access to identification of 
important communities and their properties which can be hidden
in the tail of the PageRank and hardly accessible to classification
by the PageRank algorithm.  Furthermore, the degeneracies at
various values of $\lambda$ and the
characteristic patterns directly visible in the spectra of the Google
matrix should allow to identify other hidden properties of real networks.

We thank  Leonardo Ermann and Klaus Frahm for discussions,
and CALcul en MIdi-Pyr\'en\'ees (CALMIP) for the use of their
supercomputers.

%%**********************************************************************

%\newpage

\section{APPENDIX}

Here we show the distributions of links for the AB model 
discussed in \cite{ggs} and for the university WWW network
(panel H of Fig.~\ref{fig1}).

\end{document}